\documentclass[aps,twocolumn]{revtex4}
\usepackage[utf8]{inputenc}
\usepackage{amsmath,amssymb}
\usepackage{color}
\usepackage[utf8]{inputenc}
\usepackage{amsmath,amssymb}
\usepackage{amsfonts}
\usepackage{amsmath}
\usepackage{amssymb,epsf}
\usepackage{latexsym}
\usepackage{graphicx,epsfig}
\usepackage{amssymb}
\usepackage{float}
\usepackage{subfigure}
\usepackage{epstopdf}
\usepackage[colorlinks=true,citecolor=blue,linkcolor=blue,urlcolor=black]{hyperref}

\begin{document}                                                                          %Start preparing 8 Tir 1404-Finish 10 Tir 1404%

    \title{Constraining Zero-Point Length from Gravitational Baryogenesis}
    \author{Ava Shahbazi Sooraki and Ahmad Sheykhi \footnote{asheykhi@shirazu.ac.ir}}
    \address{Department of Physics, College of
        Science, Shiraz University, Shiraz 71454, Iran\\
        Biruni Observatory, College of Science, Shiraz University, Shiraz
        71454, Iran}

\begin{abstract}
The existence of a fundamental zero-point length, $l_0$, a minimal
spacetime scale predicted by T-duality in string theory or quantum
gravity theories, modifies the entropy associated with the horizon
of spacetime. In the cosmological setup, this leads to correction
to the Friedmann equations governing the evolution of the
Universe. In this paper, we investigate the implications of
zero-point length $l_0$-corrected gravity for gravitational
baryogenesis and early universe thermodynamics, deriving
constraints on $l_0$ from observational baryon asymmetry data. We
observe that under the condition of non-equilibrium
thermodynamics, $l_0$ generates $\dot{\mathcal{R}}\neq 0$ during
radiation epoch, where $\mathcal{R}$ is the Ricci scalar. This
yields a baryon asymmetry parameter $\eta \propto l_0^2
T_D^9/M_{\rm Pl}^7$. The observed baryon asymmetry $\eta \sim 9.9
\times 10^{-11}$ constrains $l_0 \lesssim 7.1 \times 10^{-33} m$,
approximately $440$ times the Planck length. Furthermore, our
analysis reveals that the zero-point length correction in the
Friedmann equation, effectively slows the expansion rate at high
energies, resulting in a modified time-temperature relationship
where the Universe maintains higher temperatures for longer time
during early epochs compared to standard cosmology. Our results
establish zero-point length cosmology as a testable framework
connecting quantum gravity to cosmological observables, with
implications for early universe thermal history and fundamental
length scales.
\end{abstract}

\maketitle

%%%%%%%%%%%%%%%%%%%%%%%%%%%%%%%%%%%%%%%%%%%%%%%%%%%%%%%%%%%%%%%%%%%%%%%%%%%%%%%
\newpage
\section{Introduction\label{Intro}}
The Friedmann equations, which govern the expansion dynamics of
the Universe, are foundational to modern cosmology. These
equations emerge from Einstein's field equations of general
relativity (GR) under the assumption of a homogeneous and
isotropic universe described by the Friedmann-Robertson-Walker
(FRW) metric. However, at extremely high energies approaching the
Planck scale, quantum gravitational effects are expected to modify
the classical description of spacetime. One such quantum
correction arises from the concept of a fundamental zero-point
length $l_0$, a minimal length scale below which the classical
notion of spacetime breaks down. The zero-point length is
motivated by various approaches to quantum gravity, including
string theory, non-commutative geometry, and generalized
uncertainty principles (GUP). For example, in T-duality of string
theory, the momentum space propagator for massless particles is
given by~\cite{Padm,Sma,Spa,Fon,Nicolini}
\begin{eqnarray}\label{G2}
G(k)=-\frac{l_0 }{\sqrt{k^2}} K_{1}\left(l_0 \sqrt{k^2}\right),
\end{eqnarray}
where $k=p/\hbar$, $l_0$ represents the zero-point length of
spacetime, and $K_{1}(x)$ is the modified Bessel function of the
second kind. These frameworks suggest that spacetime may possess a
discrete or "fuzzy" structure at microscopic scales, leading to
modifications of physical laws near the Planck length ($l_{Pl}\sim
10^{-35} \mathrm{m}$).

Another motivation for considering zero-point length comes from
the fact that it is a possible solution to the singularity problem
in physics. It is well-known that both GR and Newton's law of
gravitation suffer from the singularity problem. Indeed, the
Newtonian gravitation potential for a spherically symmetric mass
distribution $M$ is given by  $\phi=-GM/r$, which clearly diverges
at $r=0$.  Based on relation (\ref{G2}), the static Newtonian
potential corresponding to $G(k)$ at distance $r$ was shown to
take the following form~\cite{Nicolini,Nicolini2,Nicolini3}
\begin{eqnarray}\label{phir}
\phi(r)=-\frac{G M }{\sqrt{r^2+l_0^2}},
\end{eqnarray}
where $M$ is the mass of the source of the potential and for
simplicity, we have set $\hbar=c=1$.  Thus, a point particle with
mass $m$ at distance $r=R$ from mass $M$, feels the force
\begin{equation}\label{Force}
\vec{F}=-m \vec{\nabla}
\phi(r)|_{r=R}=-\frac{GMm}{R^2}\left[1+\frac{l_0^2}{R^2}\right]^{-3/2}
\hat{r}.
\end{equation}
Based on the modified Newton's law of gravitation given in Eq.
(\ref{Force}), and applying the notion of the entropic force
scenario proposed by Verlinde \cite{Ver} and developed in
\cite{Cai2,sheyECFE}, one can derive the corrections to the
entropy associated with the horizon of spacetime \cite{KS,ShLi}.
These correction terms alter the dynamical field equations of
gravity through the thermodynamics-gravity conjecture.

While direct experimental verification of such Planck-scale
effects remains elusive, their potential implications for
cosmology and early universe physics provide a fertile ground for
testing these models. However, cosmological scenarios
incorporating a fundamental zero-point length cannot be
constrained using observational bounds from Big-Bang
nucleosynthesis (BBN), in contrast to analysis such as
Refs.~\cite{Barrow2,Anish,Ava1,Ava2}. The reason is
that the energy (and corresponding length and time) scales at
which zero-point length effects become relevant lie far above the
BBN scale, rendering primordial nucleosynthesis completely
insensitive to such corrections. Consequently, in this work we
employ gravitational baryogenesis which operates at energies many
orders of magnitude higher than BBN as a viable probe for
constraining zero-point length cosmology \cite{Luciv, Luciv2, Luciv3}.

A quintessential cosmological puzzle that may be sensitive to such
fundamental physics is the origin of the baryon asymmetry, the
observed overwhelming dominance of matter over antimatter in the
Universe \cite{ref41}. As first outlined by Sakharov \cite{ref42},
the generation of a net baryon number requires three necessary
conditions: (i) Baryon number (B) violation, (ii) C and CP
violation, and (iii) Departure from thermal equilibrium.

The Standard Model of particle physics, while containing sources
for all three conditions, is insufficient to generate the observed
asymmetry \cite{ref41}. Consequently, explanations often lie in
extensions to the Standard Model or in new physics at high
energies.

Intriguingly, the interplay between gravity and quantum mechanics
can provide a natural framework for baryogenesis. A compelling
mechanism involves a dynamical breaking of CPT (and consequently
CP) symmetry through a coupling between the baryon current $J^\mu$
and the derivative of the Ricci scalar curvature $\mathcal{R}$
\cite{ ref46}
\begin{equation}\label{intLag}
\frac{1}{M_*^2} \int d^4x \sqrt{-g}  J^\mu \partial_\mu
\mathcal{R},
\end{equation}
where $M_*$ is a cutoff scale of the effective theory and $g$ is
the determinant of the metric tensor. This interaction gives rise
to an effective chemical potential, $\mu_B \propto
-\dot{\mathcal{R}}$, for baryons versus antibaryons in the early
universe. However, in the standard cosmological framework, the
Ricci scalar vanishes ($\dot{\mathcal{R}}=0$) during the
radiation-dominated era, precisely when baryogenesis is expected
to occur, thus yielding no asymmetry.

This is where the intrinsic structure of spacetime proposed by
quantum gravity models becomes crucial. The zero-point length
$l_0$, implemented via modified horizon entropy and the resulting
corrected Friedmann equations, alters the expansion dynamics of
the Universe. This modified cosmological evolution implies that
the relation between energy density, pressure, and the Hubble
parameter is altered, leading to variations in the energy density
($\delta\rho$) and pressure ($\delta p$) from their equilibrium
values. These variations signify a deviation from thermal
equilibrium, thereby satisfying the third Sakharov condition
\cite{ref43, ref44, Luciv}.

Crucially, these $l_0$-induced perturbations result in a non-zero
$\dot{\mathcal{R}}$ during the radiation-dominated era. When
plugged into the baryogenesis mechanism of Eq. (\ref{intLag}),
this generates a calculable, non-zero baryon asymmetry parameter
$\eta$, defined as the difference between baryon and antibaryon
number densities normalized to the entropy density
\cite{Luciv,ref46}.

This approach connects a fundamental Planck-scale phenomenon
directly to an observable cosmological quantity. The measured
value of the baryon asymmetry, as constrained by the Cosmic
Microwave Background (CMB) anisotropies to be $\eta \sim (6.3 \pm
0.3) \times 10^{-10}$ \cite{Luciv,
Planck,ref48,ref49,ref50,ref51,ref52}, imposes stringent
constraints on the zero-point length parameter $l_0$.

In this paper, we explore the implications of the zero-point
length model for gravitational baryogenesis. We first derive the
modified Friedmann equations by incorporating zero-point length
corrections to the entropy of the apparent horizon. We then
investigate how this geometric incorporation of a minimal length
modifies the cosmological evolution and breaks thermal equilibrium
in the early Universe. Our primary goal is to derive the resulting
baryon asymmetry and use the observed value of $\eta$ to place
observational constraints on the fundamental zero-point length
parameter $l_0$. By doing so, we aim to bridge the gap between a
geometric implementation of a minimal length and a pivotal
cosmological observation, offering a potential pathway to test the
quantum nature of spacetime.

Our paper is organized as follows. In section II, we review the
modified cosmology based on zero-point length correction. In
section III, we present the mechanism for gravitational
baryogenesis and calculate the resulting baryon asymmetry $\eta$
within our framework and we impose the observational constraint
from the measured baryon asymmetry to establish bounds on the
zero-point length $l_0$.  In section IV, we construct the cosmic
time as a function of temperature in this cosmological framework
and disclose the effects of zero-point on time-temperature
relation. The final section is devoted to conclusions.
%%%%%%%%%%%%%%%%%%%%%%%%%%%%%%%%%%%%%%%%%%%%%%%%%%%%%%%%%%%%%%%%%%%%%%%%%%%%%%%%%%%%%%%%%%%%
\section{Modified cosmology through zero-point length correction \label{Ren}}
In this section, we review derivation of the modified Friedmann
equations inspired by zero-point length corrections to the
entropy. We follow \cite{ShLi} and apply the
thermodynamics-gravity conjecture to the apparent horizon of a FRW
universe and use the first law of thermodynamics on the apparent
horizon. It has been shown that the existence of $l_0$ alters the
entropy-area relation of black holes as well as the cosmological
horizons, which in turn affects the gravitational field equations
through the holographic principle. Taking the zero-point length
corrections into account, the entropy associated with the apparent
horizon is given by \cite{ShLi}
\begin{eqnarray} \label{St}
S_h&=& \frac{\pi R^2}{G} \left(1+\frac{l_0^2}{R^2}\right)^{-1/2}+
\frac{3\pi l_0^2}{G} \left(1+\frac{l_0^2}{R^2}\right)^{-1/2}\nonumber \\
&&-\frac{3 \pi l_0^2} {G} \ln \left(R+\sqrt{R^2+l_0^2}\right).
\end{eqnarray}
While its differential form is \cite{ShLi}
\begin{eqnarray}\label{dS1}
dS_h=\frac{2\pi R}{G} \left(1+\frac{l_0^2}{R^2}\right)^{-3/2} dR.
\end{eqnarray}
Although the associated entropy with the apparent horizon is
complicated, for $l_0^2\rightarrow 0$ it recovers the area law of
entropy, namely $S_h=\pi R^2/G$, as expected. It is evident that
$l_0$ must carry the dimension of length (or inverse energy in
natural units), since the corrected entropy-area expression must
remain dimensionless in natural units. Note that the area of the
apparent horizon is $A=4\pi R^2$, while the volume enveloped by
the apparent horizon is given by $V=4\pi R^3/3$. Note that here
$R$ is the apparent horizon radius defined as \cite{Hay1,Hay2,Bak}
\begin{equation}
\label{radius} R=\frac{1}{\sqrt{H^2+k/a^2}},
\end{equation}
where $a(t)$ is the scale factor, $H=\dot{a}/a$ is the Hubble
parameter and we assume the background spacetime is given by
\begin{equation}
ds^2=-dt^2+\frac{a^2(t)}{1-kr^2}dr^2+a^2(t)
r^2(d\theta^2+\sin^2\theta d\phi^2).
\end{equation}
The temperature associated with the apparent horizon is
\cite{Hay1,Hay2,Bak}
\begin{equation}\label{Th}
 T_h=\frac{\kappa}{2\pi}=-\frac{1}{2 \pi R}\left(1-\frac{\dot{R}}{2H R}\right),
 \end{equation}
 where $\kappa$ is the surface gravity. Let us note that for
 $\dot{R}<2H R$, one may define $T=|\kappa|/2\pi$ to avoid negative
 temperature. The energy and matter of the universe are in the form
 of a perfect fluid,
 $T_{\mu\nu}=(\rho+p)u_{\mu}u_{\nu}+pg_{\mu\nu}, $ where $\rho$ and
 $p$ are the energy density and pressure of the matter field inside
 the universe, respectively. The conservation of energy in the
 background of FRW cosmology  is guaranteed via
 \begin{equation}\label{Cont}
 \dot{\rho}+3H(\rho+p)=0.
 \end{equation}
 Since the volume of the universe increases due to its eexpansion,
 thus we have a work term in the first law of thermodynamics. The
 corresponding work density, in the background of FRW universe
 filled with a perfect fluid is obtained as \cite{Hay2}
 \begin{equation}\label{Work2}
  W=\frac{1}{2}(\rho-p).
  \end{equation}
 Taking the differential of the total energy inside the apparent
 horizon, $E=\rho (4\pi R^3/3)$, we find
 \begin{equation}
 \label{dE21}
 dE=4\pi R^{2}\rho dR-4\pi H R^{3}(\rho+p) dt,
 \end{equation}
 where we have used the continuity equation (\ref{Cont}). Combining
 Eqs. (\ref{dS1}), (\ref{Th}), (\ref{Work2}) and (\ref{dE21}) with the
first law of thermodynamics on the apparent horizon, $dE = T_h
dS_h + WdV$, after using the continuity equation (\ref{Cont}), we
arrive at
\begin{equation} \label{Fried2}-\frac{2}{R^3}\left(1+\frac{l_0^2}{R^2}\right)^{-3/2} dR=
\frac{8\pi G}{3}d\rho.
    \end{equation}
Integrating and then expanding for small values of $l_0^2/R^2$, we
reach
\begin{equation} \label{Fried3}
-\frac{2}{l_0^2}\left[ 1-
\frac{l_0^2}{2R^2}+\frac{3}{8}\frac{l_0^4}{R^4}+...\right]+C=\frac{8\pi
G}{3}\rho,
\end{equation}
where $C$ is an integration constant. The standard Friedmann
equation is restored for $l_0^2/R^2 \rightarrow 0$, provided we
choose
\begin{equation}
C\equiv \frac{2}{l_0^2}-\frac{\Lambda}{3},
\end{equation}
where $\Lambda$ is a constant which can be interpreted as the
cosmological constant. Note that one may also define $C= 2/l_0^2$
by absorbing or neglecting the contribution from vacuum energy to
the total energy density $\rho$.

Substituting $R$ from Eq. (\ref{radius}), we reach
\begin{equation} \label{Fried4}
H^2+\frac{k}{a^2}-\alpha
\left(H^2+\frac{k}{a^2}\right)^2=\frac{8\pi
G}{3}(\rho+\rho_{\Lambda}),
\end{equation}
where $\alpha=3l_0^2/4$, $\rho_{\Lambda}=\Lambda/(8\pi G)$. In the
absence of the cosmological constant ($\Lambda=0$) and for a flat
universe ($k=0$), the modified Friedmann equation is simplified as
\begin{equation} \label{Fried5}
H^2-\alpha H^4=\frac{8\pi G}{3}\rho.
\end{equation}
On the other hand, if we define $X=H^2+{k}/{a^2}$, Eq.
(\ref{Fried4}) can be rewritten as
\begin{equation} \label{Friedman41}
-\alpha X^2+ X- \frac{8\pi G}{3}\rho=0,
\end{equation}
where we have again set $\rho_{\Lambda}=0$, for simplicity. Eq.
(\ref{Friedman41}) admits the following solution
\begin{equation} \label{Friedman42}
 X=\frac{2}{3  l_0^2}\left[1\mp\sqrt{1- 8\pi G l_0^2\rho}\right].
 \end{equation}
 The standard Friedmann equation is recovered when $l_0\rightarrow
 0$, provided we choose the minus sign in the above equation. Since
 $l_0^2$ is very small, we can expand Eq. (\ref{Friedman42}). We
 find\begin{equation} \label{Friedman43}
 H^2+\frac{k}{a^2}=\frac{2}{3 l_0^2}\left[4\pi G l_0^2\rho+8\pi^2
 G^2
 l_0^4\rho^2+...\right].
 \end{equation}
The above equation may be rewritten as
\begin{equation} \label{Fried5}
H^2+\frac{k}{a^2}=\frac{8\pi G}{3}\rho\,(1+\Gamma\rho),
\end{equation}
where we have defined  $\Gamma=2\pi G l_0^2$.

Let us note that Eq. (\ref{Fried5}) in the high energy limit
(early universe) where $\Gamma\rho^2\gg \rho$ can be written as
    ($k=0$)
\begin{equation} \label{Fried55}
H^2\approx \frac{8\pi G}{3}\Gamma\rho ^2.
\end{equation}
Using $p=\rho \omega$ in the continuity equation (\ref{Cont}), we
reach
\begin{align}\label{rho}
\rho(t)= \rho{_0}\,  a^{-3(1+\omega)}.
\end{align}
Since in the early stages, our Universe is dominated by radiation,
therefore $\omega=1/3$, and $\rho_{r}=\rho_{r,0} a^{-4}$. In this
case Eq. (\ref{Fried55}) admits a solution of the form $a(t)=C_1
t^{1/4}$, with $C_1\equiv\left[ 4 \sqrt{\frac{8\pi \Gamma}{3}}
\rho_{r,0} \right]^{1/4}$. Thus in this case the scale factor
evolves as $a(t) \sim t^{1/4}$. Comparing with the radiation
dominated era in standard cosmology ($a(t) \sim t^{1/2}$), we see
that the rate of the universe expansion is slower when the
zero-point length is taken into account in the cosmological field
equations.
%%%%%%%%%%%%%%%%%%%%%%%%%%%%%%%%%%%%%%%%%%%%%%%%%%%%%%%%%%%%%%%%%%%%%%%%%%%%%%%%%%%%%%%%%%%%%%%%
\section{Baryogenesis Constraints on Zero-Point Length Cosmology \label{brz}}
This section investigates the phenomenological consequences of
zero-point length cosmology for baryogenesis, with the objective
of confronting the model with observational data and deriving
precise constraints on the fundamental scale $l_0$.

In this approach, the $l_0$-dependent corrections in the modified
Friedmann equations are treated as effective modifications to the
standard cosmological fluid. We interpret these corrections as
generating perturbations in the energy-momentum content of the
universe, representing a departure from the perfect thermal
equilibrium of standard cosmology \cite{ref43, Luciv}. To quantify
these effects, we decompose the total energy density and pressure
into their conventional equilibrium components and $l_0$-induced
perturbations as follows
\begin{align}
\rho &= \rho_0 + \delta\rho\,, \label{eq:51} \\
p &= p_0 + \delta p\,. \label{eq:52}
\end{align}
We now proceed to quantify the perturbations induced by the
zero-point length by substituting the decomposed energy density
and pressure, Eqs.~\eqref{eq:51} and \eqref{eq:52}, into the
modified Friedmann equations. The leading-order corrections are
derived as follows.

Beginning with the modified first Friedmann equation
\begin{equation}
H^{2} - \frac{3l_{0}^{2}}{4} H^{4} = \frac{8\pi G}{3}\rho.
\label{eq:53}
\end{equation}
We impose the standard general relativity limit, $H^2 = \frac{8\pi
G}{3}\rho_0$, to isolate the $l_0$-dependent contributions. This
substitution yields
\begin{equation}
\frac{8\pi G}{3}\rho_0 - \frac{3l_{0}^{2}}{4}\left(\frac{8\pi
G}{3}\rho_0\right)^2 = \frac{8\pi G}{3}(\rho_0 + \delta\rho).
\label{eq:54}
\end{equation}
Solving for the perturbation, we arrive at the leading-order
correction to the energy density
\begin{equation}
\delta\rho = -2\pi G l_{0}^{2} \rho_0^2. \label{eq:55}
\end{equation}

A parallel analysis is applied to the second modified Friedmann
equation,
\begin{equation}
\dot{H}\left(1 - \frac{3l_0^2}{2}H^2\right) = -4\pi G(\rho + p).
\label{eq:56}
\end{equation}
Inserting the decomposed expressions for energy density and
pressure from Eq.~\eqref{eq:51} into the second modified Friedmann
equation and employing the previously derived result for the
density perturbation $\delta\rho$ in Eq.~\eqref{eq:55}, the
corresponding pressure perturbation is determined. This
calculation, when performed under the conditions characterizing
the radiation-dominated epoch ($\omega=1/3$), yields
\begin{equation}
\delta p = -\frac{10\pi G l_{0}^{2}}{3} \rho_0^2. \label{eq:57}
\end{equation}
Thus, the zero-point length $l_0$ manifestly generates
non-vanishing perturbations $\delta\rho$ and $\delta p$ in the
cosmological fluid. It is evident that in the limit $l_0
\rightarrow 0$, these perturbations vanish identically, ensuring a
smooth recovery of the standard cosmological framework and
confirming the consistency of this perturbative treatment.

The perturbations in the energy density and pressure, given by
Eqs.~\eqref{eq:55} and \eqref{eq:57}, have profound implications
for early universe cosmology, particularly for the mechanism of
baryogenesis. The observed matter-antimatter asymmetry of the
universe, quantified by the baryon-to-photon ratio, stands in
stark contrast to the predictions of a universe starting from a
perfectly symmetric, thermal state \cite{ref41}. The dynamical
generation of this asymmetry, known as baryogenesis, is theorized
to occur in an expanding and cooling universe. A seminal framework
for this process was established by Sakharov \cite{ref42}, who
identified three necessary criteria: (i) violation of baryon
number ($B$) to allow for a net baryon production; (ii) violation
of both $C$ and $CP$ symmetries to ensure that the rates for
processes producing baryons and antibaryons are different; and
(iii) a departure from thermal equilibrium to prevent CPT symmetry
from undoing the generated asymmetry.

In the present model, the first two Sakharov conditions are
implemented through a well-established gravitational interaction
between the baryon current and the spacetime curvature. The
crucial third condition departure from thermal equilibrium is
naturally satisfied by $l_0$-induced perturbations $\delta\rho$
and $\delta p$ \cite{ref43, ref44, Luciv}. These perturbations,
inherent to the zero-point length corrected cosmology, represent a
fundamental departure from the perfect fluid description of the
standard radiation-dominated era, thereby providing the necessary
out-of-equilibrium environment. This approach aligns with other
quantum-gravity-inspired mechanisms, such as those employing the
Generalized Uncertainty Principle \cite{ref43} or modifications
within $f(T)$ gravity \cite{ref44}, where the breakdown of
standard thermodynamics is also central to generating the baryon
asymmetry.

The observed baryon asymmetry of the universe represents one of
the most significant outstanding problems in modern cosmology.
Gravitational baryogenesis provides a theoretically compelling
framework for explaining this matter-antimatter imbalance by
connecting it to fundamental spacetime dynamics. This mechanism,
which finds its motivation in supergravity theories and effective
field theory approaches \cite{ref43, ref45}, operates through a
dynamical breaking of CPT symmetry during cosmic expansion. The
fundamental interaction responsible for this effect couples the
baryon number current to the gradient of spacetime curvature
\cite{ref46}
\begin{equation}
\mathcal{S}_{int}=\frac{1}{M^2_*} \int d^4 x \sqrt{-g} J^\mu
\partial_\mu \mathcal{R}.
\label{eq:58}
\end{equation}
The characteristic energy scale for this interaction is set by
$M_* = (8\pi G)^{-1/2} \approx 2.4 \times 10^{18}$ GeV,
corresponding to the reduced Planck mass. This interaction
successfully addresses two of the three Sakharov conditions: the
derivative coupling to the Ricci scalar ensures the necessary C
and CP violation through dynamical CPT breaking, while baryon
number violation is assumed to be provided by additional particle
physics processes.

In the context of a homogeneous and isotropic universe, this
interaction term simplifies significantly, revealing its essential
dependence on the temporal evolution of curvature \cite{Luciv,
ref46}
\begin{equation}
\mathcal{S}_{int}=  \frac{1}{M^2_*} (n_B - n_{\bar{B}})
\dot{\mathcal{R}}, \label{eq:59}
\end{equation}
where $n_B$ and $n_{\bar{B}}$ denote the baryon and antibaryon
number densities, respectively. This expression generates an
effective chemical potential $\mu_B = -\dot{\mathcal{R}}/M^2_*$ that
distinguishes between baryons and antibaryons, with $\mu_{\bar{B}}
= -\mu_B$. The presence of this potential creates a thermodynamic
bias in the early universe's thermal plasma, preferentially
favoring baryonic states over antibaryonic ones. The resulting net
baryon number density follows from standard statistical mechanical
considerations \cite{ref46}
\begin{equation}
n_B - n_{\bar{B}} =\left|\frac{g_b}{6} \mu_B T^2\right|,
\label{eq:60}
\end{equation}
where $g_b \sim \mathcal{O}(1)$ accounts for the internal degrees
of freedom of baryonic species. This elegant formulation
establishes a direct connection between the time-varying geometry
of the universe, characterized by $\dot{R}$, and the generation of
baryon asymmetry, providing a geometric origin for one of
cosmology's most fundamental observations.

The baryon asymmetry is conventionally quantified by the
observational parameter $\eta$, defined as the net baryon number
density normalized by the entropy density \cite{ref47}
\begin{equation}
 \frac{\eta}{7} \equiv \frac{n_B - n_{\bar{B}}}{s}.
 \label{eq:61}
 \end{equation}
This parameter provides a quantitative measure of the net baryon
number density relative to the cosmic entropy density, remaining
constant throughout the adiabatic expansion of the universe.
Substituting the earlier expression for the net baryon density
from Eq.~\eqref{eq:60} and the standard thermodynamic relation for
entropy density in the radiation era, $s =
\frac{2\pi^2}{45}g_{*s}T^3$, we derive the fundamental formula for
gravitational baryogenesis \cite{Luciv, ref47}
\begin{equation}
\frac{\eta}{7} = \left|\frac{15 g_b}{4\pi^2 g_{*s}}
\frac{\dot{\mathcal{R}}}{M^2_* T}\right|.
\end{equation}
This crucial expression must be evaluated at the decoupling
temperature $T_D$, which marks the epoch when baryon number
violating interactions freeze out, thereby preserving any
generated asymmetry. The parameter $g_{*s}$ represents the
effective number of relativistic degrees of freedom contributing
to entropy; following established treatments \cite{ref47}, we
adopt $g_{*s} \approx g_* \approx 106$, corresponding to the
complete particle content of the Standard Model at high
temperatures.

The essential physical insight emerges clearly: a non-vanishing
baryon asymmetry requires $\dot{\mathcal{R}} \neq 0$. In
conventional cosmology, during radiation domination, the perfect
fluid description yields an identically zero Ricci scalar
$\mathcal{R}=0$, hence $\dot{\mathcal{R}}=0$, resulting in no
baryon asymmetry $\eta=0$. This is precisely where zero-point
length cosmology provides the crucial modification: the
$l_0$-induced perturbations to the energy density and pressure,
detailed in Eqs.~\eqref{eq:51} and \eqref{eq:52}, create the
necessary departure from perfect thermal equilibrium required by
Sakharov's third condition.

To compute the explicit form of $\dot{\mathcal{R}}$ within our
framework, we begin with the fundamental gravitational field
equation
\begin{equation}
\mathcal{R} = -8\pi G T_g, \label{eq:62}
\end{equation}
where $T_g = \rho - 3p$ is the trace of the energy-momentum
tensor. Incorporating the $l_0$-corrected density and pressure
perturbations from Eqs.~\eqref{eq:55} and \eqref{eq:57}, the trace
becomes
\begin{align}
T_g &= (\rho_0 + \delta\rho) - 3(p_0 + \delta p) \notag \\
    &= \rho_0 - 3p_0+\delta\rho - 3\delta p \notag \\
    &=\rho_0 - 3p_0 -2\pi G l_{0}^{2} \rho_0^2 + 10\pi G l_{0}^{2} \rho_0^2 \notag \\
    &=T_0+ 8\pi G l_{0}^{2} \rho_0^2.
    \label{eq:63}
\end{align}
Substituting this non-zero trace into the field equation yields the modified Ricci scalar
\begin{equation}
\mathcal{\mathcal{R}}_z =\mathcal{R}_0 -64\pi^2 G^2 l_{0}^{2}
\rho_0^2, \label{eq:65}
\end{equation}
where $\mathcal{\mathcal{R}}_0=-8\pi G(\rho_0 -3p_0)$ is the GR
Ricci scalar. For the radiation epoch ($\omega=1/3$) we have $
\mathcal{R}_0=0$ as it is expected.

Taking the time derivative and employing the continuity equation
for radiation $\dot{\rho}_0 = -4H\rho_0$ along with the standard
Friedmann relation $H^2 = \frac{8\pi G}{3}\rho_0$, we derive the
pivotal result
\begin{equation}
\dot{\mathcal{R}}_z = 512\pi^2 G^2 l_{0}^{2} H \rho_0^2 \neq 0.
\label{eq:66}
\end{equation}
The derived expression for the time evolution of the Ricci scalar
$ \dot{\mathcal{R}}_z \propto l_0^2$ , demonstrates that a nonzero
zero-point length directly drives curvature dynamics in the early
universe, with the standard cosmological result $\dot{\mathcal{R}}
= 0$ emerging smoothly in the limit $l_0 \to 0$. Consequently, the
magnitude of the generated baryon asymmetry, being linearly
proportional to $\dot{\mathcal{R}}_z$, increases with $l_0$. This
direct correlation identifies $l_0$ as the key parameter governing
the departure from thermal equilibrium and the resulting
matter-antimatter asymmetry. We now substitute Eq.~\eqref{eq:66}
into the expression for $\eta$. Using the energy density of
relativistic particles $\rho_0 ={\pi^2 g_* T^4}/{30}$ and the
Planck mass $M_*^2 = ({8\pi G}^{-1})$, a detailed calculation
yields the fundamental scaling relation
\begin{equation}
\eta = \frac{7\pi^3}{\sqrt{3} \cdot 30^{3/2}} \cdot g_*^{3/2}
\cdot \frac{l_{0}^{2} T_D^9}{M_{\rm Pl}^7}. \label{eq:68}
\end{equation}
The numerical evaluation gives the more transparent expression
\begin{equation}
 \eta \approx 0.762 \cdot g_*^{3/2} \cdot \frac{l_{0}^{2} T_D^9}{M_{\rm Pl}^7},
 \label{eq:69}
\end{equation}
where we have set the decoupling temperature $T_D = M_I \approx
3.3 \times 10^{16}$ GeV, consistent with upper bounds from
inflationary tensor modes \cite{ref43, ref46}.

\begin{figure}[H]
\includegraphics[scale=0.90]{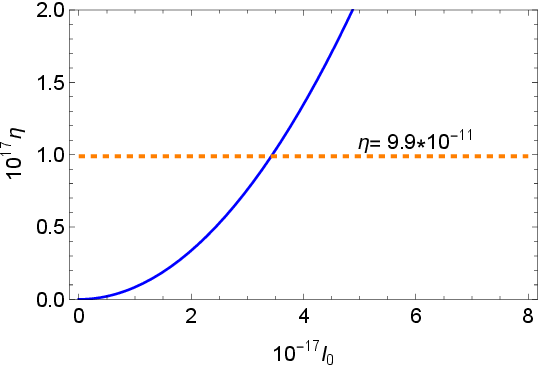}
\caption{The predicted baryon asymmetry $\eta$ as a function of
the zero-point length $l_0$. The horizontal dashed line marks the
observational upper limit. The viable parameter space, consistent
with measurements, is the area beneath this line ($\eta<9.9\times
10^{-11}$).} \label{Fig1}
\end{figure}

Confronting this prediction with the observed value $\eta_{obs}
\lesssim 9.9 \times 10^{-11}$
\cite{ref48,ref49,ref50,ref51,ref52}, we obtain a stringent
constraint on the zero-point length (see  Fig.~\ref{Fig1})
\begin{equation}
l_0 \lesssim 3.576 \times 10^{-17} \ \text{GeV}^{-1} \approx 7.1
\times 10^{-33} \ \text{m}. \label{eq:72}
\end{equation}
Expressed in natural units relative to the Planck length $l_P =
1.616 \times 10^{-35}$ m
\begin{equation}
\frac{l_0}{l_P} \lesssim 440. \label{eq:73}
\end{equation}

This bound not only validates our perturbative treatment ($l_0 \ll
1$) but also suggests that any fundamental minimal length in
nature must be proximate to the Planck scale. The characteristic
scaling $\eta \propto l_0^2 T_D^9/M_{\rm Pl}^7$ reveals the
profound sensitivity of gravitational baryogenesis to high-energy
physics and establishes a concrete connection between quantum
gravitational effects and cosmological observables.
%%%%%%%%%%%%%%%%%%%%%%%%%%%%%%%%%%%%%%%%%%%%%%%%%%%%%%%%%%%%%%%%%%%%%%%%%%%%%%%%%%%%%%%%%%%%%%%%%%%
\section{The relation between cosmic time and temperature in zero-point length cosmology \label{tT}}
This section analyzes how the zero-point length cosmology modifies
the fundamental connection between cosmic time and temperature in
the early universe. The zero-point-corrected Friedmann equations
alter the high-energy expansion dynamics and thereby lead to a
nonstandard thermodynamic evolution. As a consequence, the
time-temperature relation deviates from its familiar
general-relativistic form.

During the radiation-dominated phase, the primordial plasma
remains in thermal equilibrium, implying conservation of entropy
within any comoving volume. This standard thermodynamic
requirement leads to the condition \cite{Weinberg}
\begin{equation}
s(T)a^3 = \text{const}, \label{eq:70}
\end{equation}
where $s(T)$ represents the entropy density. The time derivative
of Eq.~\eqref{eq:70} yields the differential equation governing
entropy evolution
\begin{equation}
\dot{s}(T)a^3 + 3\dot{a}a^2 s(T) = 0 \quad \Rightarrow \quad
\frac{ds(T)}{dt}a^3 = -3\dot{a}a^2 s(T). \label{eq:71}
\end{equation}

Inserting the modified Hubble parameter in SI units
\begin{equation}
H = H_{\rm GR}\left(1 + \frac{\pi G l_0^2}{c^4}\rho\right),
\end{equation}
into Eq.~\eqref{eq:71} leads to

\begin{equation}
\frac{ds(T)}{dt} = -3\left[H_{\text{GR}}\left(1 + \frac{\pi G
l_0^2}{c^4}\rho\right)\right] s(T). \label{eq:72}
\end{equation}
This equation can be rearranged to express the differential time
element as
\begin{equation}
 dt = -\frac{ds(T)}{3s(T)}\left[H_{\text{GR}}\left(1 + \frac{\pi G l_0^2}{c^4}\rho\right)\right]^{-1}. \label{eq:73}
 \end{equation}
The resulting functional connection between cosmic time $t$ and
temperature $T$ is derived through integration of the above
expression
\begin{equation}
t = -\frac{1}{3}\int \frac{s'(T)}{s(T)}\left[H_{\text{GR}}\left(1
+ \frac{\pi G l_0^2}{c^4}\rho\right)\right]^{-1} dT, \label{eq:74}
 \end{equation}
where the prime notation indicates differentiation with respect to
temperature $T$. In the radiation-dominated regime, characterized
by $p = \rho/3$, the entropy density and energy density follow the
standard thermodynamic expressions \cite{Weinberg}
\begin{align}
s(T) &= \frac{2\mathcal{N}a_B T^3}{3}, \label{eq:75} \\
\rho(T) &= \frac{\mathcal{N}a_B T^4}{2}, \label{eq:76}
\end{align}
where $\mathcal{N}$ enumerates the total number of relativistic
degrees of freedom, accounting for all particles and antiparticles
with distinct spin states counted separately \cite{Weinberg}.
Using $\frac{s'(T)}{s(T)}=\frac{3}{T}$ in Eq.~\eqref{eq:74}, and
employing  a Taylor expansion, we obtain
\begin{equation}
t = -\int \frac{1}{TH_{\text{GR}}} \left(1 - \frac{\pi G
l_0^2}{c^4}\rho\right) dT. \label{eq:77}
\end{equation}
Substituting the expressions for $\rho(T)$ and $H_{\text{GR}} =
\sqrt{{8\pi G}/{3c^2}\rho}$ into Eq.~\eqref{eq:77} yields
 \begin{equation}
 t = -\int \left(\frac{1}{\beta T^3} - \frac{\pi G \mathcal{N} a_B l_0^2}{2c^4 \beta} T \right) dT, \label{eq:78}
\end{equation}
where we have defined $\beta \equiv \sqrt{\frac{8\pi G \mathcal{N}
a_B}{6c^2}}$. Performing the integration, we obtain
\begin{equation}
t = \frac{1}{2\beta T^2} + \frac{\pi G \mathcal{N} a_B }{4c^4
\beta} l_0^2 T^2 + \text{const}. \label{eq:79}
\end{equation}
This result can be reformulated as
\begin{equation}
t = \frac{1}{2\beta T^2}\left( 1+ \frac{3\beta^2 }{8c^2} l_0^2
T^4\right) + \text{const}. \label{eq:79}
\end{equation}
Expressing this result in terms of the zero-point length parameter
$\alpha = \frac{3l_0^2}{4c^2}$ and in SI units, we arrive at the
form
\begin{equation}
 t =\frac{1}{2\beta T^2}\left( 1+ \frac{\alpha }{2}  \beta^2 T^4\right) + \text{const}, \label{eq:80}
\end{equation}
which is the time-temperature relation in zero-point length
cosmology. In the limiting case where $l_0 \to 0$ ($\alpha = 0$),
the time-temperature relation simplifies to $t = {1}/{2\beta
T^2}$, which corresponds exactly to the standard general
relativity result \cite{Weinberg}. We note that while our modified
Friedmann equations were originally derived in natural units,
dimensional consistency necessitates that the right-hand side of
Eq.~\eqref{eq:80} must possess dimensions of time $[t]$, and
accordingly all fundamental constants have been properly restored.

\begin{figure}[H]
\includegraphics[scale=0.88]{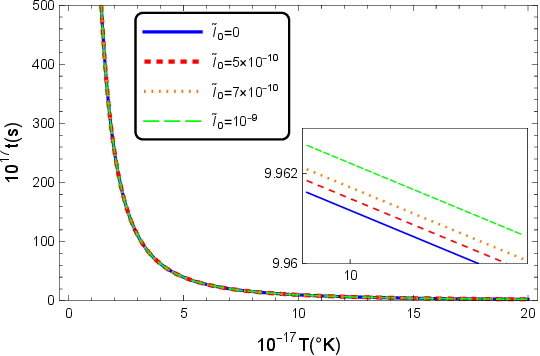}
\centering
 \caption{The behavior of the temperature vs time in the radiation dominated era
of zero-point length cosmology for different values of the
zero-point length parameter $\tilde{l}_0=[ 5\times 10^{-10},
7\times 10^{-10}, 10^{-9}]$.} \label{Fig2}
 \end{figure}

For the early universe's hot plasma composition, which includes
photons, three generations of neutrinos, antineutrinos, and
electron-positron pairs, the total effective relativistic degrees
of freedom amount to $\mathcal{N} = 43/4$. Consequently, in
centimeter-gram-second (cgs) units, Eq.~\eqref{eq:80} becomes
\begin{equation}
 t = 0.994 \left(\frac{T}{10^{10}K}\right)^{-2} F(l_0,T) + \text{const}, \label{eq:81}
 \end{equation}
 where the modification factor $F(l_0,T)$ is defined as
\begin{equation}
F(l_0,T) \equiv 1 + 4.46 \times 10^{-39} \ l_0^2 T^4.
\label{eq:82}
\end{equation}
The $l_0 \to 0$ limit eliminates all modifications ($F \to 1$),
guaranteeing consistency with standard GR \cite{Weinberg}.
Fig.~\ref{Fig2} displays the evolution of temperature $T$ versus
cosmic time $t$ across various values of the re-scaled zero-point
length parameter $\tilde{l}_0\equiv 10^{24}\; l_0$, demonstrating
that larger values of zero-point length cause higher temperatures
in the early universe.
This thermal enhancement arises from the modified expansion
dynamics: the zero-point length correction term $-{3l_0^2}H^4/4$
in the Friedmann equation effectively slows the expansion rate at
very high energies, allowing the universe to maintain higher
temperatures for longer durations compared to standard cosmology.
These results highlight the distinctive thermodynamic signatures
of zero-point length cosmology in the early universe.
%%%%%%%%%%%%%%%%%%%%%%%%%%%%%%%%%%%%%%%%%%%%%%%%%%%%%%%%%%%%%%%%%%%%%%%
\section{Conclusions} \label{Con}
The introduction of a fundamental zero-point length $l_{0}$ as a
signature of quantum gravity leads to measurable modifications in
baryogenesis predictions. By incorporating the $l_{0}$-corrected
entropy expression associated with the apparent horizon and
applying the first law of thermodynamics, we derived the modified
Friedmann equations that govern the cosmic expansion history.
These geometrically induced corrections were essential for
establishing the altered time-temperature relation in the early
universe.

In the second step, we employed the modified expansion dynamics to
compute the induced corrections to the radiation energy density
and pressure. These corrections generate a non-vanishing Ricci
scalar and its time derivative during the radiation-dominated era,
despite the fact that $\mathcal{R} = 0$ and $\dot{\mathcal{R}}=0$
in standard GR. The emergence of non-zero $\dot{\mathcal{R}}$
provides the required deviation from thermal equilibrium for
gravitational baryogenesis. The resulting baryon asymmetry scales
as
\[
    \eta \propto l_0^{\,2} T_D^{9}/M_{\rm Pl}^{7}.
\]
By comparing this theoretical prediction with the observational
value $\eta_{\rm obs} \simeq 9.9 \times 10^{-11}$, we derived a
stringent
    upper limit on the zero-point length,
    \[
    l_0 \lesssim 7.1\times 10^{-33}\,\text{m},
    \]
which corresponds to roughly $440$ times the Planck length.

We further analyzed the influence of the $l_0$-dependent term in
the modified Friedmann equation on the early-universe thermal
evolution and derived the corresponding time-temperature relation
for zero-point length cosmology. Our analysis shows that the
zero-point length alters the cosmic cooling rate, yielding $ t
={1}/{(2\beta T^2)}\left(1 + {\alpha}\beta^2 T^4/2\right). $ This
expression demonstrates that, for any fixed cosmic time, our
Universe retains a higher temperature than predicted by standard
GR. Physically, this behavior arises because the correction term
$-{3l_0^2}H^4/4$ in the modified Friedmann equation slows the
expansion rate at high energies, thereby reducing the cooling rate
and allowing the universe to remain hotter for an extended
duration. Such a modified thermal history affects several
early-universe processes, including phase transitions and
relic-abundance calculations.

The consistency between the constraint we found on $l_0$ and
expectations from quantum gravity, together with the modified
thermal history implied by the $l_0$-corrected Friedmann dynamics,
suggests that zero-point length cosmology provides a coherent and
testable framework for probing quantum-gravitational effects. The
baryon asymmetry constraint derived here constitutes a measurable
imprint of such microscopic physics and indicate that any
fundamental minimal length must be extremely small yet potentially
accessible to future precision cosmological observations.

It is important to note that, in addition to the zero-point length
correction to the area law of entropy discussed in this work,
various generalization of the Bekenstein-Hawking entropy has been
explored in the literatures \cite{Barrow, Tsallis, Kan1,Kan2}.
While the cosmological consequences of these modified entropy
(\cite{Barrow, Tsallis, Kan1,Kan2}) on the BBN at the early stages
of the Universe have been explored in
\cite{Barrow2,Anish,Ava1,Ava2}, the influences of the Kaniadakis,
Tsallis and Barrow entropies on the early baryogenesis have been
studied in \cite{Luciv,Luciv2,Luciv3}. Therefore, it is worth
constraining other modified entropies through gravitational
baryogenesis. These issues are now under investigations and the
results will be appeared elsewhere.

%%%%%%%%%%%%%%%%%%%%%%%%%%%%%%%%%%%%%%%%%%%%%%%%%%%%%%%%%%%%%%%%%%%%%%%%%%%%%%%%%%%%%%%%%

\acknowledgments{We are grateful to Shiraz university Research
Council. The work of A. Sheykhi is based upon research funded by
Iran National Science Foundation (INSF) under project No.
4022705.}
%%%%%%%%%%%%%%%%%%%%%%%%%%%%%%%%%%%%%%%%%%%%%%%%%%%%%%%%%%%%%%%%%%%%%%%%%%%%%%%%%%%%%%%%%

%We also thank the anonymous referee for very insightfuland constructive comments, which helped us improve our paper significantly.%

\end{document}